\documentclass{emulateapj}

\newcommand{\cote}{C\^{o}t\'{e}\ }
\newcommand{\jordan}{Jord\'{a}n\ }
\newcommand{\hasegan}{Ha{\c s}egan\ }
\newcommand{\etal}{et~al.\ }
\newcommand{\gz}{($g$--$z$)}

\newcommand{\hst}{{\it HST}}
\newcommand{\kmm}{{\it KMM\ }}
\newcommand{\vi}{{$V$--$I$}}

\slugcomment{Accepted for publication in the Astrophysical Journal}

\shorttitle{Color-Magnitude Relation for M87 Metal-Poor GCs}
\shortauthors{Peng et al.}

\begin{document}


\title{The Color-Magnitude Relation for Metal-Poor Globular Clusters
  in M87: Confirmation From Deep {\it HST/ACS} Imaging\altaffilmark{1}} 


\author{Eric W. Peng\altaffilmark{2,3}}

\author{Andr\'{e}s Jord\'{a}n\altaffilmark{4,5}}

\author{John P. Blakeslee\altaffilmark{6}}

\author{Steffen Mieske\altaffilmark{7}}

\author{Patrick C\^{o}t\'{e}\altaffilmark{6}}

\author{Laura Ferrarese\altaffilmark{6}}

\author{William E. Harris\altaffilmark{8}}

\author{Juan P. Madrid\altaffilmark{8}}

\author{Gerhardt R. Meurer\altaffilmark{9}}


\altaffiltext{1}{Based on observations with the NASA/ESA {\it Hubble
    Space Telescope} obtained at the Space Telescope Science Institute,
  which is operated by the Association of Universities for Research in
  Astronomy, Inc., under NASA contract NAS 5-26555.}
\altaffiltext{2}{Department of Astronomy, Peking University, Beijing
  100871, China; peng@bac.pku.edu.cn}
\altaffiltext{3}{Kavli Institute for Astronomy and Astrophysics,
  Peking University, Beijing 100871, China}
\altaffiltext{4}{Departamento de Astronom\'{\i}a y Astrof\'{\i}sica,
Pontificia Universidad Cat\'olica de Chile, Casilla 306, Santiago 22,
Chile}
\altaffiltext{5}{Harvard-Smithsonian Center for Astrophysics,
60 Garden St., Cambridge, MA 02138}
\altaffiltext{6}{Herzberg Institute of Astrophysics, 
  National Research Council of Canada, 
  5071 West Saanich Road, Victoria, BC  V9E 2E7, Canada}
\altaffiltext{7}{European Southern Observatory, Alonso de Cordova 3107,
  Vitacura, Santiago, Chile}
\altaffiltext{8}{Department of Physics and Astronomy, McMaster
  University, Hamilton, ON L8S 4M1, Canada}
\altaffiltext{9}{Department of Physics and Astronomy, Johns Hopkins
  University, Baltimore, MD 21218}


\begin{abstract}

Metal-poor globular clusters (GCs) are our local link to the earliest
epochs of 
star formation and galaxy building.  Studies of extragalactic
GC systems using deep, high-quality imaging have revealed a small but 
significant slope to the color-magnitude relation for metal-poor GCs
in a number of galaxies.  We present a study of the M87
GC system using deep, archival {\it HST/ACS} imaging with the F606W and
F814W filters, in which  
we find a significant color-magnitude relation 
for the metal-poor GCs.  The slope of this relation in the $I$ vs.\ $V$--$I$
color-magnitude diagram ($\gamma_I=-0.024\pm0.006$)
is perfectly consistent with expectations based on previously
published results using data from the ACS Virgo Cluster Survey.
The relation is driven by the most luminous GCs, those with
$M_I\lesssim-10$, and its significance is largest when fitting
metal-poor GCs brighter than $M_I=-7.8$, a luminosity which is
$\sim1$~mag fainter than 
our fitted Gaussian mean for the luminosity function (LF) of blue,
metal-poor GCs ($\sim0.8$~mag fainter than the mean for all GCs). These  
results indicate that there is a mass
scale at which the correlation begins, and is consistent with a
scenario where self-enrichment drives a mass-metallicity relationship.  We
show that previously measured half-light radii of M87 GCs from
best-fit PSF-convolved King models are consistent with the more
accurate measurements in this study, and we also explain how the
color-magnitude relation for metal-poor GCs is real and cannot be an
artifact of the photometry.  We fit Gaussian and evolved Schechter
functions to the luminosity distribution of GCs
across all colors, as well as divided into blue and red
subpopulations, finding that the blue GCs have a brighter mean
luminosity and a narrower distribution than the red GCs.
Finally, we present a catalog of astrometry and photometry for 2250 M87 GCs.

\end{abstract}




\section{Introduction}

Globular clusters (GCs) contain the oldest stellar populations in
galaxies, and are in many ways our most accessible link to the
earliest, most intense phases of galaxy building.  The formation of GC
systems are clearly linked to the formation of the galaxies that host
them.  Well-established correlations show, for instance, that both the 
mean metallicity of the entire GC population and the
mean metallicities of the well-known metallicity subpopulations
are correlated with the masses of the hosts (e.g.\ Brodie \& Huchra
1991; Larsen \etal 2001; Peng \etal 2006a).  However,
recent work has indicated that there may also be a link between the
metallicities of {\it individual} metal-poor GCs and their masses
(Harris \etal 2006; Strader \etal 2006; Mieske \etal 2006; Spitler
\etal 2006; Forte \etal 2007; Cantiello, Blakeslee \& Raimondo 2007; 
Lee \etal 2008; Wehner \etal 2008).
This discovery, mainly in the GC systems
of massive early-type galaxies, runs counter to the previously held
conventional wisdom that individual GCs do not lie on a
mass-metallicity relation.  Because of their low masses, it is not
expected that GCs would be able to self-enrich via successive
generations of star formation in the way that galaxies do.

Recently, however, high-quality color-magnitude diagrams of some
Galactic GCs and LMC star clusters show evidence of multiple stellar
populations (e.g.\ Piotto \etal 2007), provoking theoretical work on
how star clusters 
might be able to retain gas and form successive generations of stars
(D'Ercole \etal 2008; Strader \& Smith 2008; Bailin \& Harris 2009).
A mass-metallicity relation in metal-poor GCs 
is also interesting when viewed alongside the
mass-metallicity relation for dwarf spheroidal galaxies (dSphs) in the Local
Group (Kirby \etal 2008). Despite their very different structure, GCs
and dSphs have similar stellar masses.  The brightest metal-poor
GCs also appear to overlap in parameter space with objects that have been
dubbed Ultra-Compact Dwarfs (UCDs) or Dwarf-Globular Transition
Objects (DGTOs) (e.g.\ \hasegan \etal 2005, Mieske \etal 2008). How it
is that metal-poor GCs fit into the early epochs of galaxy growth, and
what their relation is to dwarf galaxies, is still unclear.

\begin{figure*}
\plottwo{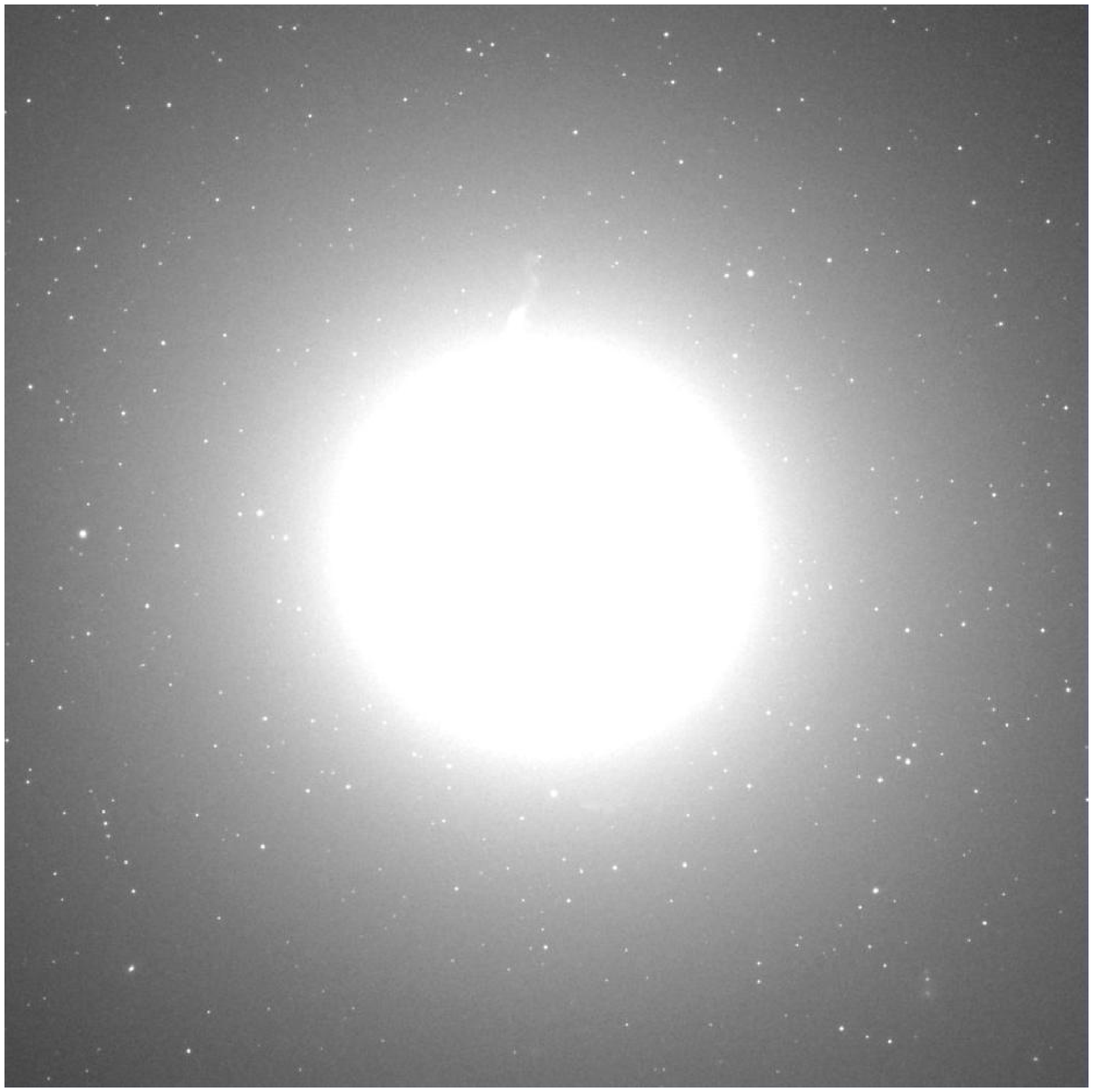}{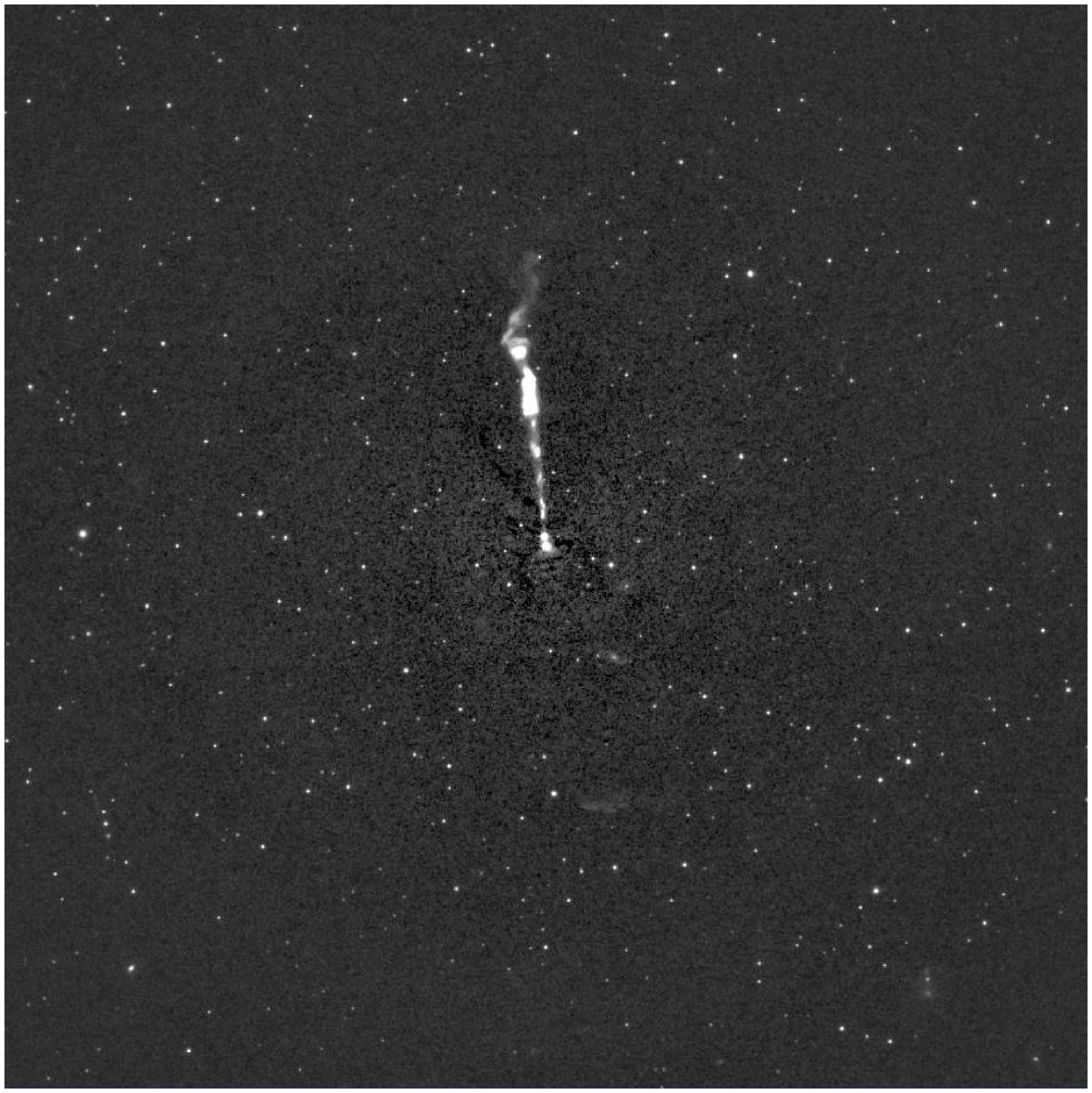}
\caption{(Left) The central $1\arcmin\times 1\arcmin$ of our reduced
  F814W image. (Right)  The same field after the subtraction of a
  smooth model.  The point sources are nearly all globular clusters.
  \label{fig:m87image}}
\end{figure*}

The question of whether or not metal-poor GCs lie on a color-magnitude 
(i.e.\ mass-metallicity relation) thus has important implications for star
cluster and galaxy formation.  However, because the 
relationship between metallicity and optical color is very steep at
low metallicities, discerning this relationship requires high quality
data, many GCs, and preferably a large wavelength baseline.
There have been many independent detections of the metal-poor GC
color-magnitude relation using a variety of data sets in a number
of different galaxies (see references above), but a recent
paper that analyzes deep archival \hst\ imaging (Waters
\etal 2009) claims not to find any significant color-magnitude relation
in the Virgo cD galaxy M87 (NGC~4486), a galaxy in which two different groups
have reported a detection with shallower HST data (Mieske \etal
2006; Strader \etal 2006 (hereafter S06)), and one group with
ground-based data (Forte \etal 2007).  Given that M87 represents possibly the
best combination of proximity and GC numbers for such a study
(McLaughlin 1999; Tamura \etal 2006; Peng \etal 2008), it
is important to establish definitively whether or not the metal-poor
GCs in M87 possess a significant relationship between color and magnitude.

In this paper, we perform an independent analysis of the same data set
as used by Waters \etal (2009).   Most of the techniques used for this paper
were first described in the analysis of the ACS Virgo Cluster Survey
(ACSVCS; \cote \etal 2004), particularly for much of the data reduction and
catalog generation (\jordan \etal 2004; \jordan \etal 2005; 
\jordan \etal 2009), and the
analysis of the color-magnitude relation of blue GCs (Mieske \etal
2006; hereafter ACSVCS~XIV).  We refer the reader to these papers for
a more thorough description of our methods.

\section{Observations and Data Reduction}

M87 was observed for 50 orbits with the ACS Wide Field Channel (WFC) in the
F606W ($V_{606}$) and F814W ($I_{814}$) filters for a microlens monitoring
program (\hst\ GO Program 10543, PI Baltz).  In all, there were 49 usable
exposures in F606W and 205 in F814W.  We processed these images using the
Apsis ACS IDT data pipeline (Blakeslee \etal\ 2003) to produce summed,
geometrically corrected, cosmic ray cleaned images for each bandpass.
Apsis measures the offsets and rotations of each individual exposure with
respect to a master catalog that it iteratively constructs from the input
images.  It then uses the measured offsets and rotations to combine the
images using the Drizzle software (Fruchter \& Hook 2002).  For this
analysis, we used the Gaussian interpolation kernel and an output image
scale of 0\farcs035~pix$^{-1}$, thus taking advantage of the very large
number of dithered exposures to improve the final resolution. 
We initially processed the
images using {\tt pixfrac}$=1.0$ and all results presented in this
paper are from this reduction.  We subsequently processed the images
using {\tt pixfrac}$=0.5$, and although the PSF
FWHM was 10\% narrower for the latter case, the differences in fitted
magnitudes and sizes were completely negligible. Total exposure times
for the combined images are 24500~s in F606W and 73800~s in F814W.   

We assume a distance to the Virgo cluster of 16.5~Mpc (Tonry \etal
2001) with a distance modulus of $31.09\pm0.03$~mag from Tonry \etal
(2001), corrected by the final results of the Key Project distances
(Freedman \etal 2001; see discussion in Mei \etal 2005b). The SBF
distance to M87 itself is consistent with the galaxy being at the
center of the Virgo cluster (Blakeslee \etal 2009).  At this distance,
one pixel in the final image therefore corresponds to 2.80~pc.  

The M87 halo light dominates the image.  In order to obtain accurate
photometry of the GC population, we first model and remove this light using
the {\tt elliprof} software developed for the SBF survey of Tonry \etal\ (1997;
see also Jord\'an \etal\ 2004).  Briefly, {\tt elliprof} iteratively fits a
series of ellipses of varying centers, ellipticities, position angles, and
Fourier component perturbations to the galaxy light distribution.  Once the
fit has converged, it interpolates to construct a smooth model, which we
subtract from the image.  Figure~\ref{fig:m87image} shows the central
$1\arcmin \times 1\arcmin$ portion of the F814W image before and after model
subtraction.

\begin{figure}
\epsscale{1.1}
\plottwo{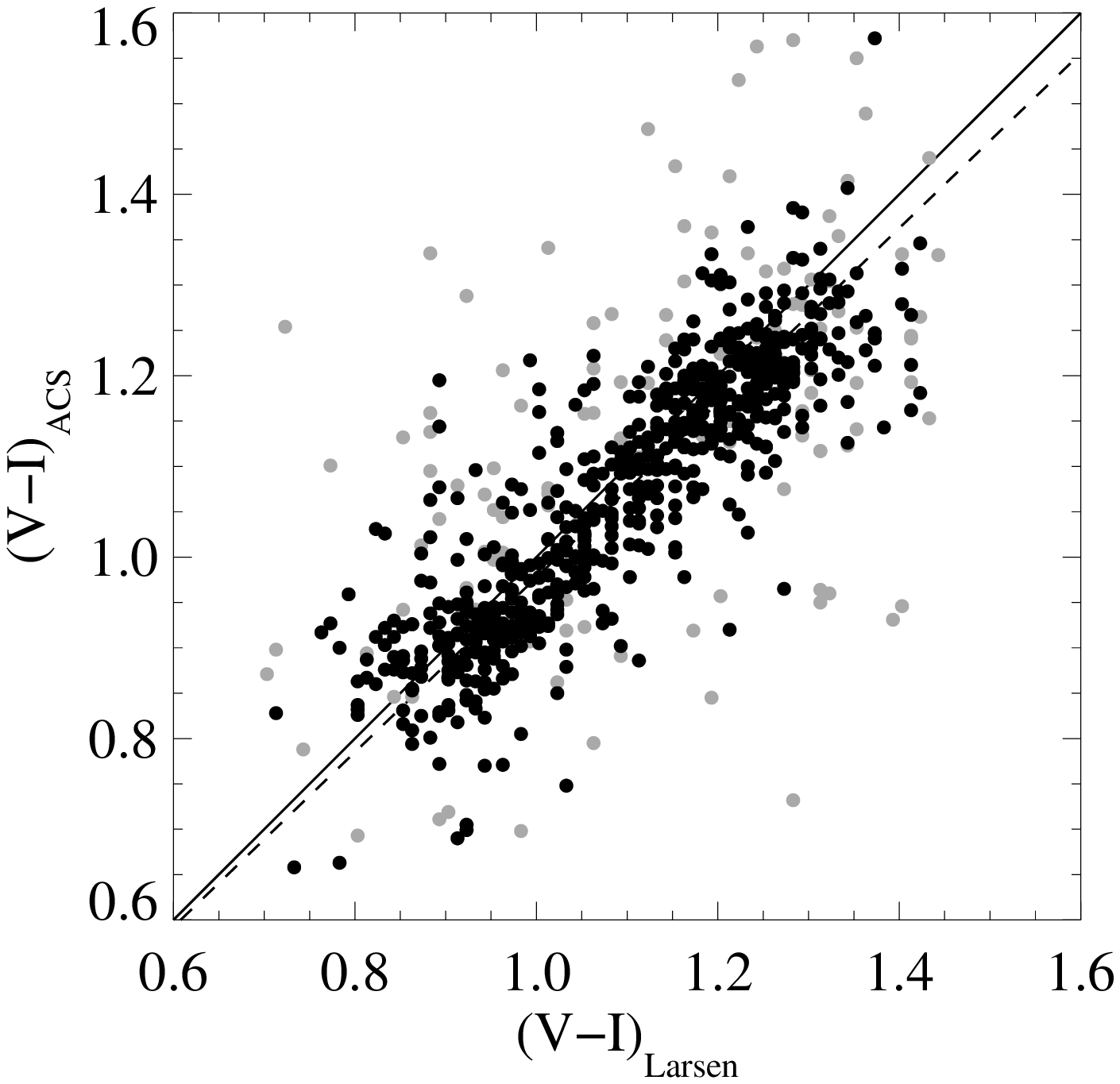}{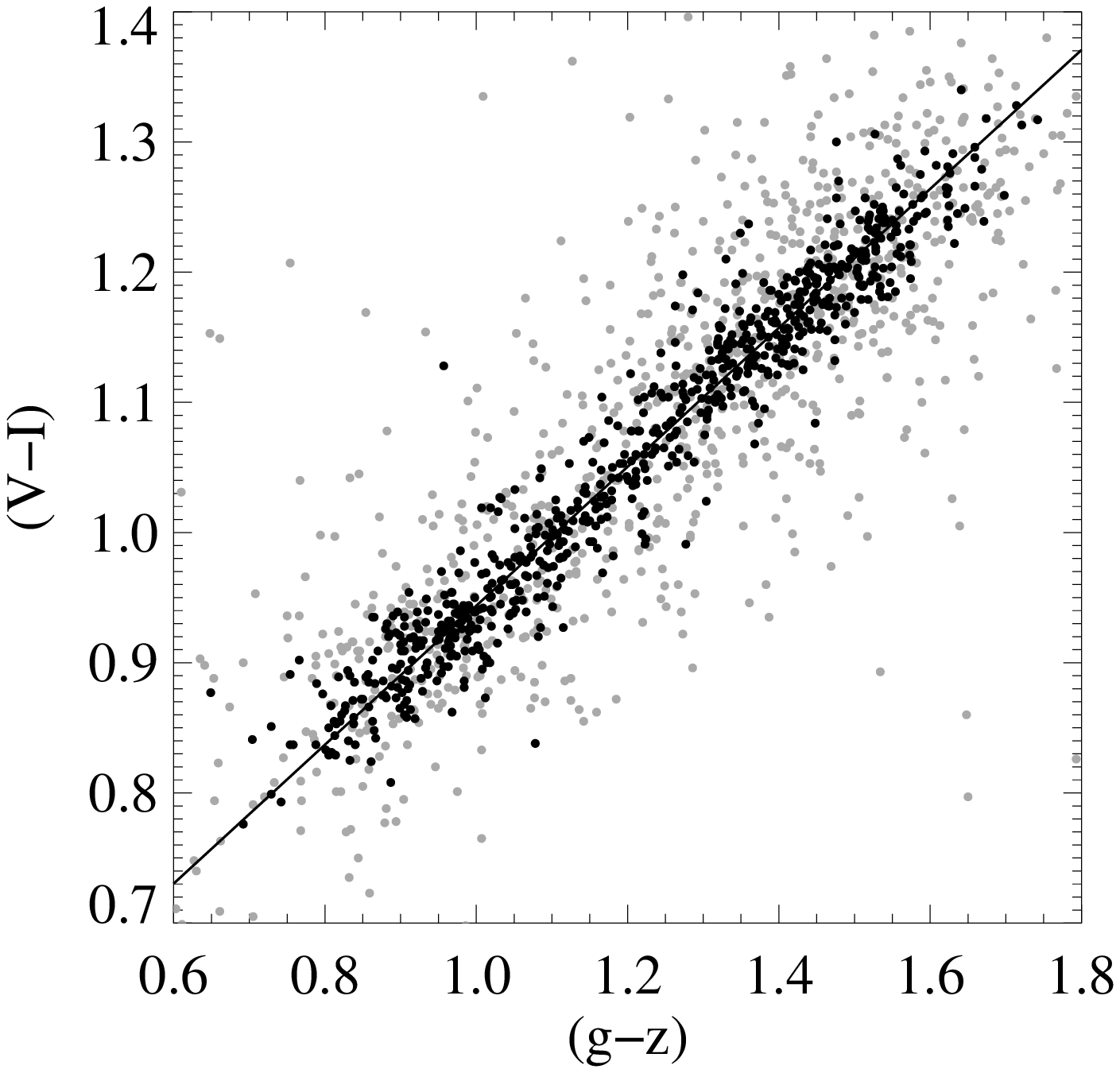}
\caption{(a, left) A comparison between our (\vi) photometry (y-axis) and the
  {\it WFPC2} (\vi) photometry of Larsen \etal (2001, x-axis). Black points
  have $\sigma_{(V-I),Larsen}<0.05$ and gray points have errors in the
  Larsen \etal photometry larger than 0.05~mag. The solid line has
  slope of unity, while the dashed line is a robust linear bisector fit to
  the black points. (b, right) A comparison
  between our (\vi) photometry and the \gz\ photometry from the
  ACSVCS.  Black and gray points are those with $\sigma_{(g-z)}$ less
  than and greater than 0.05~mag, respectively.  The line is a robust linear
  bisector fit to the black points. \label{fig:photcmp}}
\end{figure}

\section{The Photometric Catalog}

\subsection{Detection and Photometry}
\label{sec:phot}

Detection was performed with SExtractor (Bertin \& Arnouts 1996) on the
model subtracted images using a procedure similar to that described in
Jordan \etal (2004; ACSVCS~II), including the removal of large-scale model
residuals with SExtractor as described there. Object detection on the
final subtracted image was performed using a detection threshold of
five connected pixels at a $10\sigma$ significance level. The depth of
the data is such that this high threshold still allows the detection
of the great majority of GCs present in the field of view, while at
the same time avoiding the detection of SBF and other real, albeit much
fainter features. The detections in both the F606W and F814W images
were finally matched using a matching radius of $0\farcs1$.

Our catalog of matched sources was culled by removing objects which
have a mean elongation $\epsilon \equiv a/b$, measured in the F606W
and F814W filters to be $\langle \epsilon\rangle \ge 2$, as well as
those with magnitudes more then 7~mag from the expected GCLF turnover.  This
procedure removes 47 objects, which are visibly background galaxies. We
note the objects removed tend to be faint, with only two rejected
objects with $I \lesssim 21$. Our catalog has 2459 GC candidates after
rejecting elongated sources. The sources left at this stage are run
through KINGPHOT (Jordan \etal 2005), a code that measures structural
and photometric parameters by fitting the two-dimensional ACS surface
brightness profiles with PSF-convolved isotropic, single-mass King
(1966) models. 

PSFs for the F606W and F814W bands were obtained by
drizzling the position-dependent set of PSFs constructed by Anderson \&
King (2006). The Anderson \& King (2006) PSFs are meant to be used on
the geometrically distorted ACS chips, so they should be drizzled in
the same fashion as the actual data in order to be used on re-sampled,
drizzled frames. Reproducing the kernel and {\tt pixfrac} used is important
to determine accurate sizes. In order to measure the structural
parameters, we used a fitting radius of 6 pixels ($0\farcs21$).

\begin{figure}
\epsscale{1}
\plotone{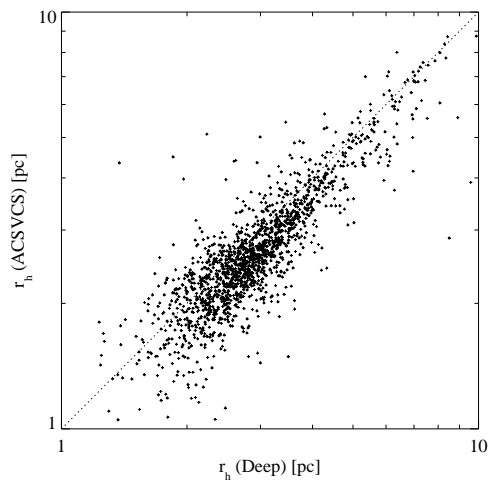}
\caption{A comparison between half-light radii ($r_h$) measured in the
  single-orbit ACSVCS data and $r_h$ measured in the deep F606W and
  F814W ACS images used in this study.  Measurements of the same
  objects in these two independent data sets agree very well within
  the expected errors, showing that the ACSVCS data provides the
  necessary depth and resolution to measure GC sizes at Virgo
  distance.  The dotted line represents a slope of unity. The slight
  offset from unity of $0.26$~pc, which equals 3 milliarcseconds or
  less than one tenth of a pixel, is
  within the expected error from PSF systematics.
  \label{fig:rhrh}}
\end{figure}

For each object classified as a globular cluster candidate, KINGPHOT
is used to measure the total magnitude, King concentration index, $c$,
and half-light radius, $r_h$ in both bandpasses. Note that these
magnitudes already include the effect of size. As described in Jordan
\etal (2009; ACSVCS Catalogs Paper), the magnitudes still require an
aperture correction due to the fact that the PSFs used to fit the data
do not extend far enough to include all of the light. We used a method
identical to the one described in Jordan \etal (2009) to obtain aperture
corrections using mean PSFs constructed up to a radius of $3\arcsec$
and used in Sirianni et al (2005) to derive aperture
corrections. Because the magnitudes derived by KINGPHOT already
include the effects of the different size of each source, the aperture
corrections ($A$) are roughly constant with $r_h$, and have values of
$A_{\rm F606W} \approx 0.07$ mag and $A_{\rm F814} \approx 0.06$ mag
(we use the measured $r_h$ for each object, averaged on both bands, to
apply the aperture correction that applies to that size). Magnitudes
were de-reddened using a value of $E(B-V)=0.023$ obtained from the the
DIRBE maps of Schlegel et al (1998) and extinction ratios appropriate
for a G2 star as presented in Sirianni et al (2005). 

To produce our final sample of GCs, we rejected all objects with
magnitude errors greater than 0.5~mag in either band (116 objects),
or with measured half-light radii larger than 10~pc (140 objects).
Nearly all bona fide GCs, which have a mean half-light
radius of 2.7~pc (\jordan \etal 2005), will qualify under this size
criterion. 

In order to make sure, however, that this cut does not
affect our results, we used KINGPHOT to specially fit a subset of the
extended objects.  Because of our intial choice of fitting radius,
sources with 
$r_h>10$~pc have unreliable measurements (see \jordan \etal 2005 for a
description of the biases that may arise when the fitting radius is
less than $\sim$2 times $r_h$). We selected the 24 extended objects with
$I<22$~mag and $0.7<(V-I)<1.5$~mag 
for re-fitting with KINGPHOT using a fitting radius of 18~pixels
(50~pc).  Of these 24 objects, 5 have very large sizes and are clearly
background galaxies, and one was identified spectroscopically as a
background galaxy by \hasegan \etal (2005).  Of the remaining 18, the five 
most luminous, extended objects are in fact 
ultra-compact dwarfs (UCDs) or dwarf-globular transition objects
(DGTOs) identified 
by \hasegan \etal (2005).  Whether or not they should be included in a
sample of GCs is debatable, but including or excluding them does not
affect the results of this study.  The 116 
remaining fainter extended objects (likely background galaxies) are
not in an important magnitude or color range for this study, and were
excluded from our analysis.  Our final GC catalog contains 2250 sources.

Instrumental magnitudes were converted to VEGAMAG magnitudes in the
F606W and F814W filters using zeropoints of 26.398 and 25.501 mag,
respectively (Sirianni \etal 2005). Conversion of these magnitudes to
$V$ and $I$ is not altogether advisable, as this conversion depends on the
spectrum of the sources, but is necessary for comparison purposes. We
converted magnitudes to $V$ and $I$ by using the relation between $(V-I)$ and
($m_{\rm F606W} - m_{\rm F814W}$) presented in Equation (3) of
DeGraaff et al (2007), namely $(V-I) = 1.2(m_{\rm F606W} -
m_{\rm F814W}) + 0.06$, and the following equation from Table~22 in
Sirianni et al., $I=m_{\rm F814W, OBMAG} + 25.495 - 0.002(V-I)$ (see
discussion in DeGraaff \etal as to why our adopted color transformation
is preferred to either the observed or synthetic transformations
presented in Sirianni et al; the relation between $I$ and $m_{\rm
F814W}$ is less problematic, with the observed and synthetic
transformations presented in Sirianni \etal being almost identical).

The full table of 2250 GC positions and KINGPHOT $V$ and $I$ photometry is
presented in Table~\ref{tab:gctable}.  We set the zeropoint of the astrometry
to the International Celestial Reference Frame (ICRF; Fey \etal
2004).  We did this by first matching 1579 matched GCs with the ACSVCS
catalog and determining the median offsets between the two catalogs in
RA and Dec, both of which were less than $1\arcsec$.  The dispersion
about the median offset between 
the two catalogs was very small ($\sigma=0.012\arcsec$).  We then
calibrate the astrometry directly to the ICRF by measuring the
position of the nucleus of M87 in the ACSVCS F475W image, and compare
it to the position measured by Fey \etal (2004) using very long
baseline interferometry radio observations.

\begin{deluxetable}{ccccccc}
\tablewidth{0pt}
\tablecaption{Positions and KINGPHOT photometry for M87 GCs}
\tablehead{
\colhead{ID} & 
\colhead{RA(J2000)} & \colhead{Dec(J2000)} &
\colhead{$V$} & \colhead{$\sigma_V$} &
\colhead{$I$} & \colhead{$\sigma_I$} \\
\colhead{} & 
\colhead{} & \colhead{} &
\multicolumn{4}{c}{(mag)}
}
\startdata
    1 &   187.7432637 &    12.4052854 &  25.793 &  0.034 &  24.574 &  0.051 \\
    2 &   187.7371764 &    12.3819827 &  23.703 &  0.012 &  22.696 &  0.018 \\
    3 &   187.7341900 &    12.3709317 &  23.293 &  0.008 &  22.342 &  0.014 \\
    4 &   187.7331924 &    12.3676597 &  22.327 &  0.019 &  21.255 &  0.022 \\
    5 &   187.7340066 &    12.3707719 &  23.047 &  0.009 &  21.927 &  0.014 \\
\enddata
\label{tab:gctable}

\tablecomments{The complete version of this table is in the electronic
edition of the Journal.}
\end{deluxetable}


\subsection{Photometric and Size Comparisons}
\label{sec:photcomp}

Because the quality of the photometry is one of the core issues for
this analysis, we present comparisons between this photometry and
those from two previous studies.  First, we compare to the $V$ and $I$
photometry from the deep {\it HST/WFPC2} study of Larsen \etal (2001).
This study used the F555W and F814W filters and transformed to $V$ and
$I$.  We matched 757 objects between the two catalogs that were
within a radius of $0\farcs2$.  The WFPC2 field of view is smaller
than that of ACS/WFC, and contains fewer objects overall. 
Figure~\ref{fig:photcmp}a shows the
comparison between the two photometric catalogs.  We fit the objects
with low error ($\sigma_{(V-I)_{Larsen}}<0.05$~mag) with a robust
linear bisector to obtain $(V-I)_{ACS} = 0.015+0.963(V-I)_{Larsen}$, a
relation that very nearly has a slope of unity, showing that our
photometry is consistent with that from the Larsen \etal (2001) study.

\begin{figure}
\plotone{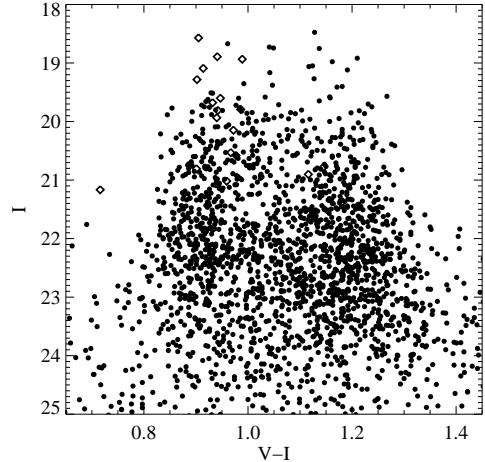}
\caption{The $V$--$I$ color-magnitude diagram of 2250 M87 globular
  clusters (dots). Overplotted as diamonds are 18 objects with $r_h>10$~pc,
  some of which are confirmed UCDs or DGTOs (\hasegan \etal 2005).
  Excluding these more 
  extended objects from our analysis has a no impact on our
  results. \label{fig:m87cmd}}
\end{figure}

Next, we compare to the ACS F475W and F850LP (hereafter $g$ and $z$)
photometry from the ACSVCS.  Figure~\ref{fig:photcmp}b shows this
comparison using a matched catalog between the two data sets created with a
matching radius of $0\farcs1$.  The (\vi) and \gz\ colors are
well-correlated, and a robust linear bisector fit to the high
signal-to-noise data ($\sigma_{(g-z)}<0.05$) produces the relation
$(V-I) = 0.944 + 0.534\{(g-z)-1\}$ (normalized to $(g-z)=1$ for
convenience).  This is nearly identical to the relation fit by Peng
\etal (2006b).  Again, the shallower and deeper data are consistent
within the internal uncertainties of both data sets.

\begin{figure*}
\centerline{\plottwo{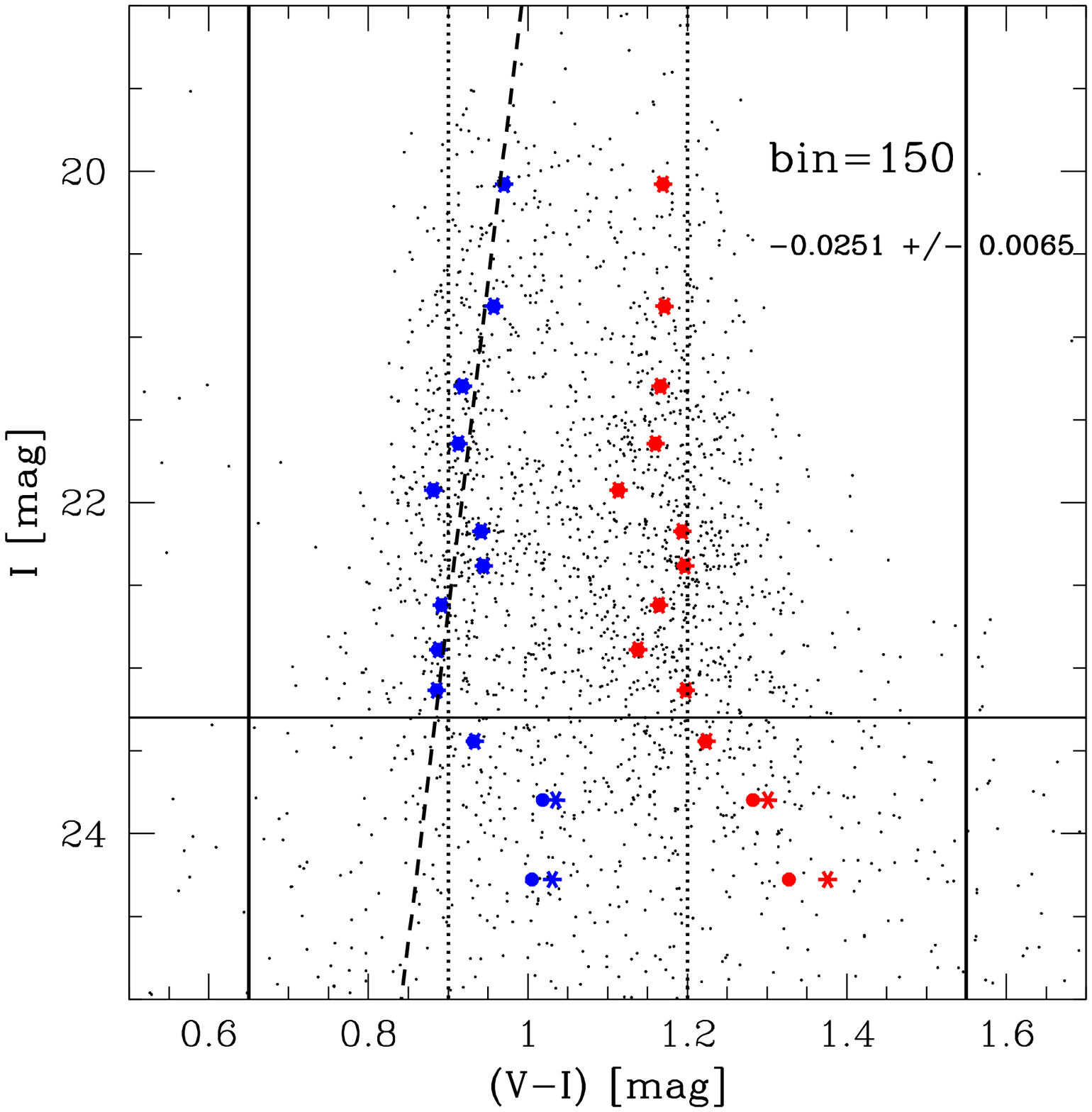}{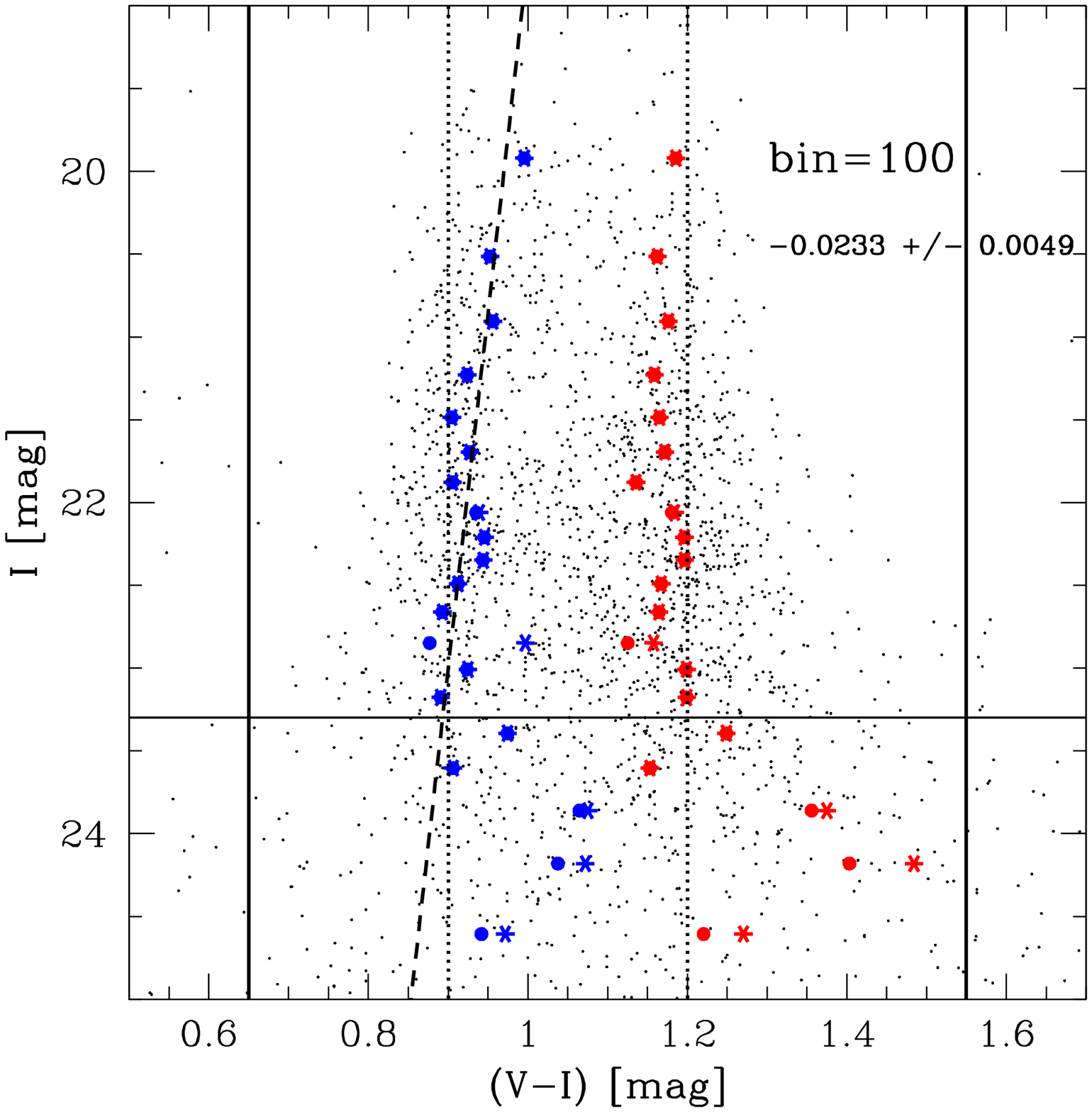}}
\caption{\label{fig:cmdfit} 
  Color-magnitude diagram of M87's GC system from the deep HST imaging
  set. KMM fit results are indicated, analogous to Figures~1 and 2 of
  ACSVCS~XIV. Left panel corresponds to a luminosity bin size of
  N=150, right panel to N=100. We fit a linear relation to the KMM
  blue peak position. The dashed lines indicate fit to data points
  with $-12< M_I < -7.8$ mag ($19<I<23.3$ mag, horizontal solid line). The two
  vertical solid lines indicate the blue and red color limit we 
  adopted for fitting the KMM peak positions.  The corresponding
  slope values are given in the plot legend, and also in
  Table~\ref{tab:table}.} 
\end{figure*}

Lastly, we compare the measured half-light radii, $r_h$ between the
ACSVCS data and the much higher signal-to-noise ACS data in this
study.  The ability to measure the sizes of the GCs is not only
scientifically interesting in its own right, but is critical for
photometry because the magnitudes are affected by the derived $r_h$.
Figure~\ref{fig:rhrh} shows the comparison between $r_h$ measured
independently in the two data sets.  We find that there is
excellent correlation and agreement between half-light radii measured
in the shallower ACSVCS data and those measured in the current deep
ACS data for the same objects.  There is a slight offset of median
0.26~pc between the measured radii, which is consistent with
the expected systematic uncertainty from the different PSFs used in the
different filters.  This offset is equal to $0\farcs003$ on the sky,
or less than 0.1 pixels on the ACS/WFC detector.  Both Kundu (2008) and Waters
\etal (2009) claim that the single-orbit ACSVCS data might not be deep
enough for an accurate measure of GC size, but \jordan\ \etal (2005) 
already used simulations to show that accurate sizes can be measured in the
ACSVCS data, and we now confirm this empirically using the
deeper data in the present study. 

\section{The GC Color-Magnitude Diagram and Fitted Relations}
\subsection{Fitting the Color-Magnitude Relations}

In Figure~\ref{fig:m87cmd} we present the (\vi) M87 globular
cluster color-magnitude diagram (CMD).  The two color subpopulations,
blue and red, are 
clearly visible, with a division around (\vi)$\approx 1.05$.  The
typical error in (\vi) for GCs with $I<23.5$, is $\sigma_{(V-I)}<0.02$~mag.
We have also overplotted the 18 extended objects with $r_h>10$~pc to
show where they lie in the CMD.  The brightest five objects with
colors similar to the metal-poor GCs are UCD/DGTOs identified by
\hasegan \etal (2005).  The following analysis does not
include the extended objects, although we have performed the analysis
on the combined sample and find virtually no difference in the results.

We analyzed the
luminosity dependence of the GC colors in this CMD in an identical
fashion as in ACSVCS~XIV. In brief, we apply the heteroscedastic mode
of \kmm to the CMD, subdivided into luminosity bins containing the
same number of data points. We use two different bin sizes of N=100
and N=150 to quantify how much the result depends on the specific
binning chosen. The fitted mean positions of the blue and red peaks
are plotted over the respective CMDs in Figure~\ref{fig:cmdfit}, as a
function of magnitude. Apart from varying the bin size, we also choose
two different pairs of initial guesses for the blue and red peak:
(\vi)$=0.9$ and 1.15, and (\vi)$=1.0$ and 1.25. Finally, we adopted
limiting magnitudes of $-12<M_I<-7.8$ mag ($19.0<I<23.3$~mag) for the
fitting. We fit linear relations to the \kmm 
peak positions for all pairs of initial guesses, bin sizes and
limiting magnitudes. 

For each bin size, the slope $\gamma_I=\frac{d(V-I)}{dI}$
is adopted as the mean of the value derived from the two pairs of
initial guesses. The errors of the fit are derived from re-sampling
the points using the observed scatter around the fitted relation
for the dispersion. The
difference between the slopes derived from peak positions of the two
different initial guesses was negligible compared to the formal fit
errors. In Table~\ref{tab:table} we give the resulting values of the slope
$\gamma_I$ for both the blue and red
peak, for the two bin sizes, and for the average of the two.
In the end, the fitted value for the slope is robust to the exact
choice of bin size.  For the remainder of this paper, we use the
average $\gamma_I$ of the fits in the two bin sizes, giving
$\gamma_{I,\rm blue}=-0.024\pm0.006$, and 
$\gamma_{I,\rm red}=+0.003\pm0.007$. 

\begin{table}
\begin{center}
\small\vspace{0.2cm}
\caption{Color-magnitude trends for blue and red globular clusters belonging to 
M87, determined with \kmm fits\label{tab:table}}
\vspace{0.2cm}
\begin{tabular}{|l||c|c|}
\hline Sample &  $\gamma_{I,\rm blue}$ &  $\gamma_{I,\rm red}$\\
\hline
N=100, $M_{I,faint}=-7.8$ & $-0.0232 \pm 0.0049$ & $+0.0033\pm 0.0052$\\
N=150, $M_{I,faint}=-7.8$ & $-0.0251 \pm 0.0065$ & $+0.0026\pm 0.0079$\\
\hline
{\bf Average} & {\bf $-0.0241\pm 0.0057$} & {\bf $+0.0030 \pm 0.0066$} \\
\hline

\end{tabular}
\vspace{0.2cm}\\ Notes: Columns 2 and 3 give the slopes
$\gamma_I=\frac{d(V-I)}{dI}$ of the blue and red GC
subpopulations, as derived from linear fits to \kmm determined peak
positions (see text and caption of Figure~\ref{fig:cmdfit}).  Errors are
from random resampling of the data points using their measured
dispersion around the fit.\normalsize
\end{center}
\end{table}

To check the effect of excluding extended, UCD-like objects, we also
perform the above analysis including the 18 extended sources.  Doing
so only changes $\gamma_{I,\rm blue}$ by 0.001, which is much smaller
than the uncertainty.  We have also checked the color-magnitude
diagram using the 
SExtractor {\tt mag\_auto} aperture photometry to make sure that
the color-magnitude relation is not affected by our GC selection in general.
Of the objects rejected by the size criterion, nearly all are either too
faint ($I>22.5$) or too red ($(V-I)>1.1$) to impact our analysis of
the metal-poor GCs.
We have performed our analysis 
including all sources that make our initial SExtractor detection
threshold, and find that doing so does not change our conclusions. 

We emphasize that these observations are extremely deep, and that the
scatter in the color-magnitude diagram is caused by physical
variations in color, not by photometric error.  Therefore, the uncertainty
in the measurement of the color-magnitude relation is intrinsic to the
population of GCs, and cannot be improved without more GCs.

\subsection{The Slope Dependence on Limiting Magnitudes}

We also investigate this dependence of the slope on limiting
magnitudes in order to find the luminosity range over which the
correlation is most prevalent. There have been previous suggestions
that the slope in the blue GCs is more strongly defined by the more
luminous GCs (Harris \etal 2006; Harris 2009). We quantify this more
directly by performing our analysis using a range of faint and bright
limiting magnitudes.  

\begin{figure}
\plotone{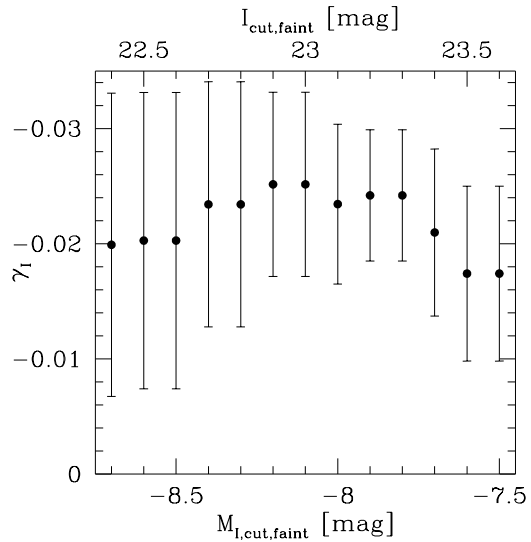}
\caption{\label{fig:slopesfaint} 
  We show the fitted slope ($\gamma_I$) as a function of the faint end
  limiting magnitude of the metal-poor GCS used in the fit
  ($M_{I,cut,faint}$).  We find 
  that the color-magnitude relation has a similar slope for all
  samples down to $M_{I,cut,faint}\approx-7.8$ and has maximum
  significance at this 
  limit.  Including GCs fainter than this limit, or approximately
  0.5~mag fainter than the turnover of the GCLF, has the effect of
  beginning to wash the trend out.}
\end{figure}

In Figure~\ref{fig:slopesfaint}, we fit the blue GCs in the
same way as described above, but varying the faint magnitude cut
within the range $-8.7<M_I<-7.5$, in steps of 0.1~mag.  These fits
show that the slope is consistently at the same value when fitting
only the brightest GCs and is at maximum significance 
when the blue GC faint limit is $M_I<-7.8$.  This corresponds to
a luminosity 1.1~mag fainter than the Gaussian mean of the GC
luminosity function for blue, metal-poor GCs, and 0.8~mag fainter than
the GCLF turnover for all GCs (see Appendix~\ref{sec:gclf}).

In Figure~\ref{fig:slopesbright}, we fit the blue GCs in the
same way as described above, but varying the {\it bright} magnitude cut
within the range $-12.1<M_I<-9.0$, in steps of 0.2~mag.  In this
figure, the fitted slope rapidly declines as the most luminous GCs are
excluded from the fit.  It is clear that slope is driven by the bright
blue GCs with $M_I\lesssim -10$.

However the sample is fitted, we cannot avoid the conclusion that the
metal-poor GCs in the bright half of the GCLF possess a significant
color-magnitude relation.  This relation is driven by the most
luminous blue GCs, but continues to rise in significance with the
inclusion of GCs down to $M_I=-7.8$, about 1~mag below the Gaussian
mean of the blue GCLF. The metal-rich GCs possess no significant slope.

\section{Discussion}
\subsection{Comparisons to Previous Estimates}

\begin{figure}
\plotone{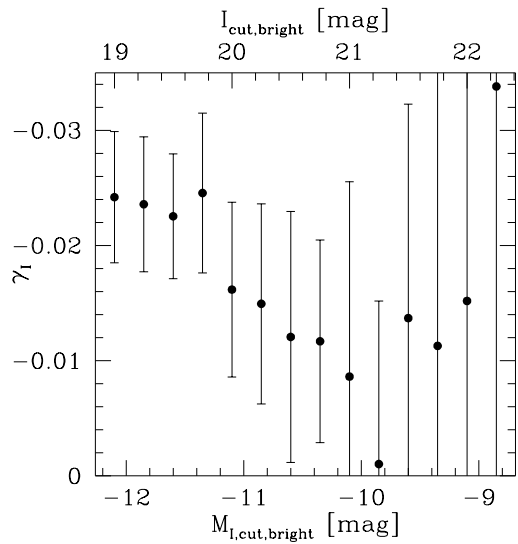}
\caption{\label{fig:slopesbright} 
  We show the fitted slope ($\gamma_I$) as a function of the bright
  end limiting magnitude of the metal-poor GCS used in the fit
  ($M_{I,cut,bright}$).  We find that the color-magnitude
  relation is driven by the most luminous GCs, or those with
  $M_I\lesssim -10$.  For this plot, we assumed
  $M_{I,cut,faint}=-7.8$.}
\end{figure}

There have now been many estimates of the slope of the metal-poor
GC color-magnitude relation, but they have often been done in different
filter systems with different instruments.  More importantly, these
slopes are usually transformed into a mass-metallicity relation
using a metallicity-color relation derived from either Galactic
GCs or stellar population synthesis models.  Although transforming to
physical quantities is important for understanding the ultimate origin
of this phenomenon, we stress that for comparisons between different
studies it is the slopes in {\it observed quantities} (color
and magnitude) that should be compared rather than in the transformed
quantities (metallicity and mass).  This is for the simple reason
that transformations between different filter sets and colors are much
more precisely known than the transformation from color to
metallicity.  Direct comparisons in mass-metallicity space that do not take
this into account can produce misleading results, as we will describe
below for the case of Waters \etal (2009).

The fits in ACSVCS~XIV used a limiting magnitude of $M_z=-7.7$, which
closely matches a magnitude limit of 
$M_I=-7.8$ (the $gz$ photometry is on the AB system while $VI$ is
VEGAMAG).  Thus, for the purposes of our comparison, we will use the
average of the measured slope values fitted to samples with a limiting
magnitude of $M_I=-7.8$, which is $\gamma_I=-0.022$.  In ACSVCS~XIV,
the measured slope for
M87's metal-poor GCs is $\gamma_z=\frac{d(g-z)}{dz} =
-0.042\pm0.015$.  This is virtually identical to that measured by S06, 
$\gamma_z=-0.043\pm0.010$.  Assuming that the slope
measured in ACSVCS~XIV is correct, what would be the expected slope in
the $V$ and $I$ bandpasses used for this study?  We have already above
in Section~\ref{sec:photcomp} matched 1637 objects in common between
our current catalog and the ACSVCS catalog for M87.  The fit between
\vi\ and $g-z$ presented above for these objects yields a slope of 0.534.
Likewise, $I = z - 0.342 + 0.225(g-z)$.  Therefore, to transform from
$\gamma_z$ to $\gamma_I$ we use the relation:
\begin{equation}
\gamma_I = \gamma_z \frac{d(V-I)}{d(g-z)} \frac{dz}{dI}
         = \gamma_z \times \frac{0.534}{1+0.225\gamma_z}
\end{equation}
For $\gamma_z = -0.042\pm0.015$ from ACSVCS~XIV, we therefore expect $\gamma_I =
-0.022\pm0.008$.  This value is indistinguishable from the slope we derived in
the previous section, $\gamma_I = -0.024\pm0.006$, and thus the
color-magnitude relation for the 
original, single-orbit, ACSVCS $g$ and $z$ imaging is perfectly
consistent with that derived from the deep, 50-orbit $V_{606}$ and
$I_{814}$ observations analyzed here.

\subsection{$m_{606}-m_{814}$ versus $g-z$: Depth, Wavelength
  Baseline, and the Color-Magnitude Relation}

The advantage of the ACS data analyzed in this paper and in Waters
\etal (2009) is the unparalleled depth compared to other imaging of
M87.  However, for the purpose of detecting a mass-metallicity
relation in the metal-poor GCs, this data has the singular
disadvantage of a short wavelength baseline.  Because the filters are
F606W and F814W, the baseline is shorter than even the traditional $V$
and $I$ to which the instrumental colors are transformed.  Given the
transformations used in this paper $(m_{606}-m_{814}) \propto
0.445(g-z)$.  Thus, the errors in ($m_{606}-m_{814}$) need to be 0.445
times smaller than in \gz\ to achieve the same metallicity
sensitivity, corresponding to a higher $S/N$ by a factor of 2.25.
In the case of the observations used here, the median ratio between
the error in ($m_{606}$--$m_{814}$) and the ACSVCS error in \gz\ for
objects in common
is 0.229.  These observations are thus in principle between 1.7 and 2
times more sensitive to metallicity than the ACSVCS observations until
scatter in the color become dominated by systematic error at the
bright end, or by the intrinsic scatter in color among GCs (the latter
is the case for our observations).  Although the
ratio of \hst\ orbits is 50:1, the short 
wavelength baseline of the filters used and the intrinsic scatter of
color among GCs explains why these observations are not nearly as
much of an improvement over the ACSVCS as one might initially expect.

\subsection{The Discrepancy with Waters \etal (2009)}

The main conclusion of Waters \etal (2009) was that they did not
detect a color-magnitude relation for the metal-poor GCs in M87 using
deeper imaging than had previously been analyzed.  In this paper,
however, we have performed an independent analysis and reduction of
the same archival data, and we find a color-magnitude relation with
the exact slope expected from the results of ACSVCS~XIV and
S06.  In this section, we make three comments on the
claims, results, and methodology adopted by Waters \etal (2009).


\subsubsection{The Color-Magnitude Relation of Blue GCs is not
    an Observational Artifact.}  

Both Waters \etal (2009) and Kundu (2008) make the
argument that the color-magnitude relation previously detected for blue
GCs is due to improperly performing standard aperture photometry
of resolved GCs.  The basic
claim is that if metal-poor GCs have a size-luminosity relation
where more luminous GCs are larger, then applying a single aperture
correction to all objects leads to biases in color that correlate with
magnitude.  This claim is can be disproven in three ways.  

First, both this study and the
ACSVCS~XIV study of the GC color-magnitude relation use photometry that
{\it always explicitly take into account the size of the GC}.  As
explained above, and in \jordan \etal (2005, 2009), we use model
magnitudes derived from PSF-convolved King model fits.  

Second, Kundu (2008)
and Waters \etal (2009) further claim that the ACSVCS data is not
deep enough to measure the sizes of GCs at Virgo distance.  As
has been shown using simulations (\jordan \etal 2005) and now comparing
to the deep ACS data analyzed in this paper, reliable sizes and photometry
can in fact be derived from the shallower ACSVCS data used as the basis
for previous detections of the blue GC color-magnitude relation.
Moreover, the scatter in the size-luminosity relation for GCs is much
more important than its slope, which is shallow and only
relevant for the highest luminosity GCs.

Third, a simple test shows that even if one were to perform standard
aperture photometry on resolved GCs of different sizes, the
resulting bias in {\it color} is much too small to create the
observed color-magnitude 
relation.  \jordan \etal (2009) performed simple 4~pixel radius
aperture photometry on PSF-convolved King models with a range of 
sizes.  They found that although the total flux of GCs can be significantly
underestimated for $r_h\!>\!3$~pc, the color of the GC is only minimally
affected.  With $1\!<\!r_h\!<\!30$~pc, the \gz\ aperture correction deviates
by only $^{+0.004}_{-0.014}$ from the $r_h=3$~pc fiducial.  For this
kind of simple photometry, the different sizes of the PSFs in F475W and
F850LP indeed cause the measured \gz\ color to become increasingly
biased to the red up to a maximum of 0.004~mag at $r_h\sim 7$~pc.
However, at larger sizes the bias {\it reverses direction} and the
measured GC colors start becoming biased to the blue.  This can be
explained because for larger sizes, we enter the regime where the
size difference between the PSFs in the two filters are small
compared to the size of the GC (10~pc $\sim0\farcs125$ at Virgo).
Therefore, even for simple 4-pixel aperture photometry, there is no
conspiracy between size and magnitude that can artificially produce
the color-magnitude relation in the blue GCs.  This explains how
slopes determined from different studies --- those accounting for GC size
(ACSVCS~XIV, this paper), or simply applying an average aperture
correction (S06), or with galaxies at different distances (Harris \etal
2006), or with data from the ground where GCs are unresolved
(e.g. Forte \etal 2007; Wehner \etal 2008) --- all find similar
results. (See also a discussion on aperture corrections in Harris 2009;
Harris \etal 2009).

\subsubsection{The Color-Magnitude Diagram of Waters \etal (2009) shows
    Significant Scatter.}  
Despite the high signal-to-noise of
these deep ACS F606W and F814W observations, the M87 GC
color-magnitude diagram presented in Figure~2 of Waters \etal (2009)
appears to contain a large amount of scatter.  This is in contrast
to the CMD presented in this paper which uses the same data
(Figure~\ref{fig:m87cmd}), as well as the \gz\ CMDs in 
Figure~1 of ACSVCS~XIV, and in Figure~4 of S06.  
In particular, we draw the reader's attention to the many bright
blue sources in the Waters \etal CMD, those with $I<22$ and $(V-I)<
0.9$, of which there are at least 46 shown.  By
contrast, our catalog generated from the same imaging contains only
2 objects (6 before the various cuts described in
Section~\ref{sec:phot}) in the equivalent region ($I<22$ and
$(V-I)<0.8$, because 
we adopt the DeGraaff \etal (2007) transformation for $V$ and $I$ as
opposed to the one from Sirianni \etal (2005), resulting in a
$\sim0.1$~mag shift at these colors).  We do not know the cause for
this difference, 
but we note that none of the recently published GC CMDs for M87,
including Waters \etal (2006), show evidence for sources that are so
luminous and blue.  The same kind of disagreement is seen for luminous
red sources, with $(V-I)>1.3$.  Given the short wavelength baseline of this
filter set, any increase in the photometric scatter would make it
extremely difficult to detect a color-magnitude relation
in the blue GCs of $\gamma_I\approx0.02$.  

In addition, we note that Waters \etal (2009) use the flux
within a 4-pixel aperture to perform their concentration-dependent aperture
corrections.  Thus, the amount of raw information they use to derive their
photometry for each object is no different from that used by KINGPHOT
except that their photometry is not from direct King model
fits.  Their analysis, therefore, does not include any additional flux
or information as compared to ours.

\subsubsection{The Importance of Comparing Color-Magnitude Relations
    Instead of Mass-Metallicity Relations.}  
There are many transformations between GC color and metallicity in the
literature, based on different data sets.  Most studies either derive an
empirical relation from the known colors and metallicities of
Galactic globular clusters, or they use models of simple stellar
populations.  What has become clear over the years is that the
relationship between optical color and metallicity for old stellar
populations is nonlinear, and becomes increasingly steep for 
metal-poor objects ([Fe/H]$\lesssim-1$) (Cohen \etal 2003, Peng \etal
2006a, Cantiello \& Blakeslee 2007).  This is exactly the color and
metallicity regime with which we
are concerned in this paper and in all studies of metal-poor GCs.

Although it is necessary and even desirable to transform
color-magnitude relations into mass-metallicity relations, it is
important that observational comparisons be done {\it directly},
rather than through the minefield of different metallicity-color
relations.  Waters \etal (2009) do not compare colors directly, but
instead attempt
to simulate slopes in their data starting with mass-metallicity
relations.  
The [Fe/H]--(\vi)
relationship assumed by Waters \etal (2009), however, is inconsistent with the
[Fe/H]--\gz\ relations assumed in ACSVCS~XIV and S06; (\vi) is less
sensitive than to [Fe/H] than \gz, particularly at low metallicity. This leads
Waters \etal to conclude that the expected slope would be much more
easily seen in their data than it actually is.  Furthermore, their
simulated CMD takes the brighter GCs as fixed and offsets GCs with
$I>20$ to the blue. If they had instead fixed the GCs around the
GCLF turnover and then offset brighter ($I<23$) GCs to the red the
expected trend would be more difficult to detect.  As a result, 
they generate a large gap between the blue and red GC populations
that, in their Figure~3a, bears no resemblance to a real CMD.  
If previous authors had observed such a CMD to derive
a color-magnitude relation, there would be little argument about its
existence.  The simulations of Mieske \etal (2006), which were
performed in color space alone, are better suited to testing the
sensitivity of observations to any color-magnitude relation.

\subsection{Substructure in the Color-Magnitude Diagram}

Having established the existence of a color-magnitude relation for
metal-poor GCs, it is still not clear what its underlying cause might
be.  The high photometric accuracy of this data set, however, shows
some tantalizing hints at substructure in the color-magnitude diagram
shown in Figure~\ref{fig:m87cmd}, especially in the blue GCs.  While
there is an obvious ridge of blue GCs that appears to be driving the
color-magnitude relation for $21<I<23$, there are also clusterings of
GCs in the CMD at higher luminosities and intermediate colors, as well
as ``void'' around $(V-I)\approx 1.0$ and $I\approx 21.5$.  Given
the precariousness of transforming from broadband color to metallicity
for stellar populations with [Fe/H]$\lesssim-1$ (e.g., Peng \etal
2006a; Yoon \etal 2006; Cantiello \& Blakeslee 2007), we do not wish
to over-interpret 
these groupings.  It is possible, however, that this clustering, which
appear to strengthen the observed color-magnitude relation, may be the
sign of an intermediate population of GCs.  


\section{Conclusions}

We present an analysis of the color-magnitude diagram and luminosity
functions for globular
clusters in the Virgo cD galaxy, M87, using  deep, archival 
{\it HST/ACS} imaging  in the F606W and F814W filters.  We report an
independent detection at high significance ($4\sigma$) of a color-magnitude
relation for the blue GCs, which was previously reported for this galaxy
using shallower data from the ACSVCS (Mieske \etal 2006; S06).  The
measured slope in (\vi) is entirely consistent with the previously
published values in \gz.  This finding is contrary to a recent
independent reduction and analysis of the same deep archival data by
Waters \etal (2009) who claim to find no relation.

We fit the color-magnitude relation for a range of faint and bright limiting
magnitudes, $M_I$, to test the idea that the bright GCs drive the
relation.  We find that the slope is driven by GCs brighter than
$M_I\approx -10$, and is most significant for samples
including GCs brighter than $M_I=-7.8$, or 1~mag fainter than the mean
of the blue GCLF, or 0.8~mag fainter than
the mean of the total GCLF.  This suggests that there is a mass scale at
which the correlation between mass and metallicity begins, and is
qualitatively consistent with a scenario where self-enrichment drives
the relation (Bailin \& Harris 2009).

All of our photometry is performed using the King model fitter,
KINGPHOT (\jordan \etal 2005), and explicitly takes into account the
size of each object.  We show that the half-light radii previously
measured using KINGPHOT on the shallower ACSVCS data in \jordan \etal
(2005) are well-correlated with the new, more accurate measurements.

We explain that the color-magnitude relation seen in the metal-poor GCs
cannot be an observational artifact involving aperture corrections, as
argued by Kundu (2008) and Waters \etal (2009).  All of our photometry
explicitly uses the fitted size of the object to derive total
magnitudes and colors.  We also show that even if one were to use a
fixed aperture correction for all GCs (as in S06 and Harris \etal
2006), the magnitude of the bias is much too small to create the
observed color-magnitude relations.

\acknowledgments
We thank S{\o}ren Larsen for sharing his {\it HST/WFPC2} photometry of
M87 GCs.  E.~W.~P.\ gratefully acknowledges the support of the Peking
University Hundred Talent Fund (985).  A.~J. acknowledges support from
the Chilean Center of Excellence in Astrophysics and Associated
Technologies, and from the Chilean Center for Astrophysics FONDAP
15010003. This research has made use of the NASA/IPAC Extragalactic
Database (NED) which is operated by the Jet Propulsion Laboratory,
California Institute of Technology, under contract with the National
Aeronautics and Space Administration. 

Facilities: \facility{HST(ACS,WFPC2)}



\appendix
\section{The GC Luminosity Functions of M87 GC Populations}
\label{sec:gclf}

\begin{figure}
\plotone{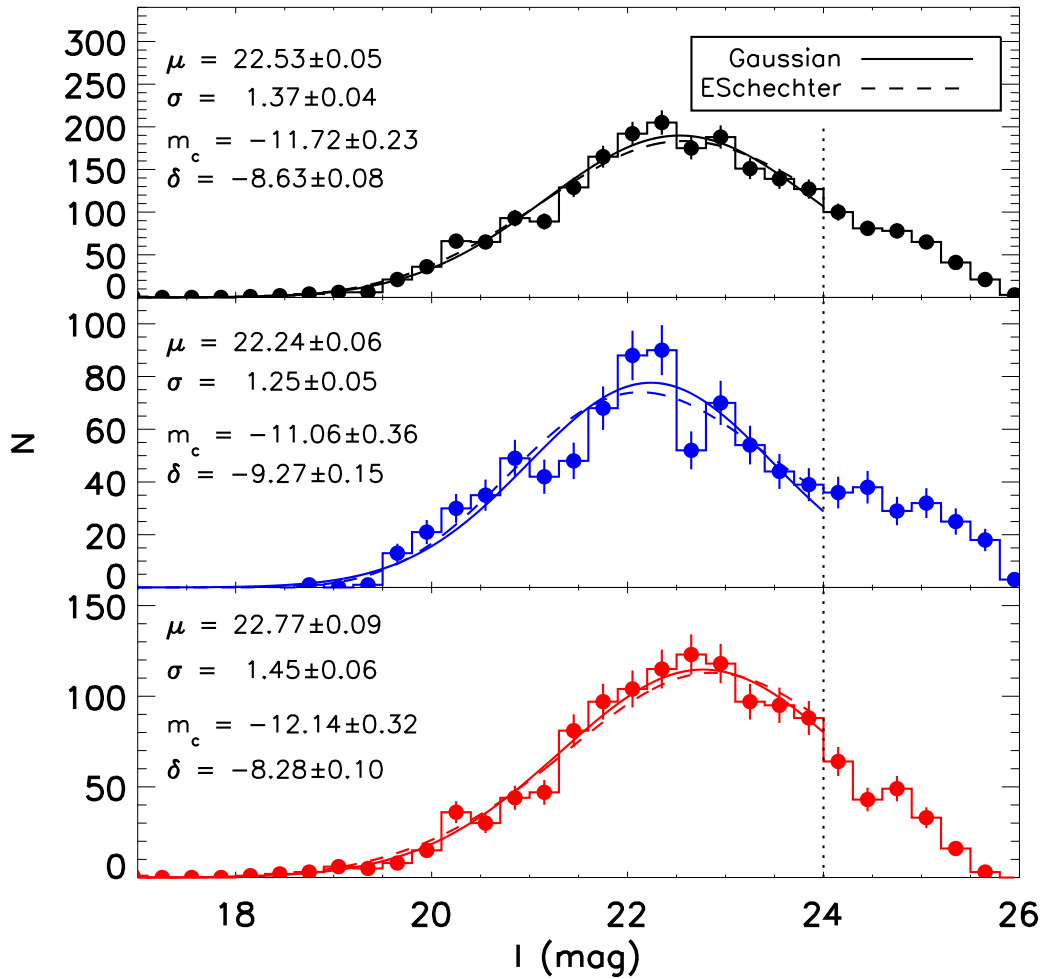}
\caption{The luminosity functions of M87 GCs, divided by color.  We
  plot the luminosity functions of all observed GCs (top), blue GCs with
  $(V-I)<1.04$ (middle) and red GCs with $(V-I)>1.04$.  We performed
  parametric fits to the GCs with $I<24$ (marked by dotted line) using
  both Gaussian (solid curve) and evolved Schechter (dashed curve)
  functional forms.  The best-fit parameters for each function are
  listed in the plots and in Table~\ref{tab:fits}.  The blue GCs have
  a brighter peak and a narrower width than do the red GCs.
  \label{fig:gclf}}
\end{figure}

The quality and completeness of our photometry also allows us to
fit the luminosity function of the M87 GCs.  The color-magnitude
diagram in Figure~\ref{fig:m87cmd} hints that the luminosity functions
of the blue and red GC subpopulations have different means and
widths.  We quantify this by fitting the $I$-band luminosity functions of the
GC subpopulations with two different functions: a Gaussian and an
evolved Schechter function (see \jordan\ \etal 2007). We fit
only GCs with $I<24$~mag, a limit above which neither contamination nor
completeness is a problem, but which is still $\gtrsim 1\sigma$
fainter than the mean in a Gaussian parametrization.  We divide
blue from red GCs at $(V-I)=1.04$, the color at which a GC is equally
likely to belong either to the blue or red subpopulation according to
a double Gaussian homoscedastic fit using the Kaye's Mixture Model
(KMM; McLachlan \& Basford 1988; Ashman, Bird, \& Zepf 1994).  The
results of our fits are presented in Table~\ref{tab:fits}, and the
luminosity distributions with fits are shown in
Figure~\ref{fig:gclf}.  We note that although we have plotted the GC
magnitude distributions in bins, the fits were performed on the
unbinned data using maximum likelihood estimation.  There appears to
be an excess of faint objects at $I>24$~mag which are likely to be
compact background galaxies.

The best-fit Gaussian parameters of the total population are
$\mu_I=22.53\pm0.05$ and $\sigma=1.37\pm0.04$.  This is nearly
identical to the measurement by Kundu \etal (1999) of
$\mu_{I,K99}=22.55\pm0.06$ and $\sigma_{I,K99}=1.41\pm0.11$. Using a
distance to M87 of 16.5~Mpc ($m-M=31.09$), our measurement translates to 
$\mu_{M_I}= -8.56\pm0.05$~mag. 

The Gaussian mean of the blue GC subpopulation is expected to be
brighter than that of the red GC subpopulaton if the GC mass function
is universal across metallicity (e.g., Ashman, Conti, \& Zepf 1995;
Puzia \etal 1999; \jordan \etal 2002; \jordan \etal 2007).  When fitting the
individual subpopulations, we find that  
$\mu_{I,blue}=22.24\pm0.06$~mag and $\mu_{I,red}=22.77\pm0.09$~mag (or
$\mu_{M_I}=-8.85$~mag and $-8.32$~mag, respectively).
This difference, $\mu_{I,blue}-\mu_{I,red}=-0.43$~mag, is consistent
with what Puzia \etal (1999) find for the GC system of M49.

The Gaussian widths of the GCLFs for the two subpopulations are also
different.  We measure $\sigma_{I,blue} = 1.25\pm0.05$~mag and
$\sigma_{I,red} = 1.45\pm0.06$~mag.  The LF of the blue GCs is
narrower than that for the red GCs, but is not as narrow as the mean
widths for GCLFs of early-type dwarf galaxies in the Virgo and
Fornax clusters, which can have $\sigma\sim1.0$~mag (\jordan \etal
2006, 2007: Miller \& Lotz 2007).

We also fit an evolved Schechter function, as defined by \jordan \etal
(2007), equation 8.  We fix the faint end slope of the cluster initial
mass function to have a power law
exponent $\beta=2$, as was done in \jordan \etal (2007).  The
evolved Schechter function then has two free parameters, which we
represent as $m_c$, the 
absolute magnitude of the exponential ``cutoff'' associated with the
bright end of the Schechter function, and $\delta$, the absolute
magnitude representing
the average mass loss per cluster over a Hubble time (for details, see
\jordan \etal 2007, Section 3.2).  The fitted parameters for all GCs
(with $I<24$~mag) are $m_{c,I}=-11.72\pm0.23$~mag and
$\delta_I=-8.63\pm0.08$~mag.  For the blue and red subpopulations, we
find that $m_{c,I,blue}=-11.06\pm0.36$~mag, $m_{c,I,red}=-12.14\pm0.32$, 
$\delta_{I,blue}=-9.27\pm0.15$, and $\delta_{I,red}=-8.28\pm0.10$. 
{}

\clearpage



\begin{table}
\begin{center}
\small\vspace{0.2cm}
\caption{Best-fit parameters for $I$-band GC Luminosity
  Functions\label{tab:fits}}
\vspace{0.3cm}
\begin{tabular}{|l||ccc||cc|}
\hline Sample & \multicolumn{3}{|c||}{Gaussian} &
\multicolumn{2}{|c|}{Evolved Schechter}\\
 & $\mu_{I}$ & $\mu_{M_I}$ & $\sigma_{I,\rm all}$ & $m_{c,I}$ & $\delta_{I}$ \\
\hline
All & $22.53\pm0.05$ & $-8.56\pm0.05$ & $1.37\pm0.04$ & $-11.72\pm0.23$ & $-8.63\pm0.08$ \\
Blue & $22.24\pm0.06$ &$-8.85\pm0.06$& $1.25\pm0.05$ & $-11.06\pm0.36$ & $-9.27\pm0.15$ \\
Red & $22.77\pm0.09$ &$-8.32\pm0.09$& $1.45\pm0.06$ & $-12.14\pm0.32$ & $-8.28\pm0.10$ \\
\hline
\end{tabular}
\vspace{0.2cm}\\ Notes: All parameters have units of magnitudes.  Blue
and red GCs are divided at $(V-I)=1.04$~mag. Fits are performed on GCs
with $I<24$~mag.  Absolute magnitudes assume $m-M=31.09$.\normalsize
\end{center}
\end{table}

















\clearpage









\begin{thebibliography}{}


\bibitem[Anderson 
\& King(2006)]{2006acs..rept....1A} Anderson, J., \& King, I.~R.\ 2006, Instrument Science Report ACS 2006-01, 34 pages, 1 

\bibitem[Ashman et al.(1994)]{1994AJ....108.2348A} Ashman, K.~M., Bird, 
C.~M., \& Zepf, S.~E.\ 1994, \aj, 108, 2348

\bibitem[Ashman et al.(1995)]{1995AJ....110.1164A} Ashman, K.~M., Conti, 
A., \& Zepf, S.~E.\ 1995, \aj, 110, 1164 

%
%
%
%

\bibitem[Bailin 
\& Harris(2009)]{2009arXiv0901.2302B} Bailin, J., \& Harris, W.~E.\ 2009, arXiv:0901.2302 


%
%
%
%

\bibitem[Bertin \& Arnouts(1996)]{1996A&AS..117..393B} Bertin, E., \&
Arnouts, S.\ 1996, \aaps, 117, 393

%
%
%
%
%
\bibitem[Blakeslee et al.(2003)]{2003ASPC..295..257B} Blakeslee, J.~P., 
Anderson, K.~R., Meurer, G.~R., Ben{\'{\i}}tez, N., \& Magee, D.\ 2003, 
Astronomical Data Analysis Software and Systems XII, 295, 257

\bibitem[Blakeslee et al.(2009)]{2009ApJ...694..556B} Blakeslee, J.~P., et 
al.\ 2009, \apj, 694, 556 

%
%
\bibitem[Brodie \& Huchra(1991)]{1991ApJ...379..157B} Brodie, J.~P., \& 
Huchra, J.~P.\ 1991, \apj, 379, 157

%
%
\bibitem[Cantiello et al.(2007)]{2007ApJ...668..209C} Cantiello, M., Blakeslee, J.~P., \& Raimondo, G.\ 2007, \apj, 668, 209

\bibitem[Cantiello \& Blakeslee(2007)]{2007ApJ...669..982C} Cantiello, M., 
\& Blakeslee, J.~P.\ 2007, \apj, 669, 982

%
%
%
%
%
%
%
%
%
\bibitem[C{\^ o}t{\' e} et al.(2004)]{2004ApJS..153..223C} C{\^ o}t{\' e}, 
P., et al.\ 2004, \apjs, 153, 223

%
%
%

\bibitem[D'Ercole et al.(2008)]{2008MNRAS.391..825D} D'Ercole, A., 
Vesperini, E., D'Antona, F., McMillan, S.~L.~W., 
\& Recchi, S.\ 2008, \mnras, 391, 825

%
%
%
\bibitem[DeGraaff et al.(2007)]{2007ApJ...671.1624D} DeGraaff, R.~B., 
Blakeslee, J.~P., Meurer, G.~R., \& Putman, M.~E.\ 2007, \apj, 671, 1624 

%
%
%
%
%
%
%
%
%
%
%
%
%
%
%

\bibitem[Forte et al.(2007)]{2007MNRAS.382.1947F} Forte, J.~C., Faifer, F., 
\& Geisler, D.\ 2007, \mnras, 382, 1947 

\bibitem[Freedman et al.(2001)]{2001ApJ...553...47F} Freedman, W.~L., et 
al.\ 2001, \apj, 553, 47

\bibitem[Fruchter 
\& Hook(2002)]{2002PASP..114..144F} Fruchter, A.~S., \& Hook, R.~N.\ 2002, \pasp, 114, 144 

%
%
%
%
%
%
%
%
%
%
%

\bibitem[Harris et al.(2006)]{2006ApJ...636...90H} Harris, W.~E., Whitmore, 
B.~C., Karakla, D., Oko{\'n}, W., Baum, W.~A., Hanes, D.~A., 
\& Kavelaars, J.~J.\ 2006, \apj, 636, 90 

\bibitem[Harris (2009)]{2009ApJ...submitted} Harris, W.~E.,
  2009, \apj, submitted

\bibitem[Harris et al.(2009)]{2009AJ....137.3314H} Harris, W.~E., 
Kavelaars, J.~J., Hanes, D.~A., Pritchet, C.~J., 
\& Baum, W.~A.\ 2009, \aj, 137, 3314 

\bibitem[Ha{\c s}egan et al.(2005)]{2005ApJ...627..203H} Ha{\c s}egan,
  M., et al.\ 2005, \apj, 627, 203

%

\bibitem[Jord{\'a}n et al.(2002)]{2002ApJ...576L.113J} Jord{\'a}n, A., 
C{\^o}t{\'e}, P., West, M.~J., \& Marzke, R.~O.\ 2002, \apjl, 576, L113 

%
\bibitem[Jord{\' a}n et al.(2004)]{2004ApJS..154..509J} Jord{\' a}n, A., et 
al.\ 2004, \apjs, 154, 509



\bibitem[Jord{\'a}n et al.(2005)]{2005ApJ...634.1002J} Jord{\'a}n, A., et 
al.\ 2005, \apj, 634, 1002 

\bibitem[Jord{\'a}n et al.(2006)]{2006ApJ...651L..25J} Jord{\'a}n, A., et 
al.\ 2006, \apjl, 651, L25

\bibitem[Jord{\'a}n et al.(2007)]{2007ApJS..171..101J} Jord{\'a}n, A., et 
al.\ 2007, \apjs, 171, 101 (Paper XII)

\bibitem[Jord{\'a}n et al.(2009)]{2009ApJS..180...54J} Jord{\'a}n, A., et 
al.\ 2009, \apjs, 180, 54 


%
%
\bibitem[King(1966)]{1966AJ.....71...64K} King, I.~R.\ 1966, \aj, 71, 64
%
%
%

\bibitem[Kundu et al.(1999)]{1999ApJ...513..733K} Kundu, A., Whitmore, 
B.~C., Sparks, W.~B., Macchetto, F.~D., Zepf, S.~E., 
\& Ashman, K.~M.\ 1999, \apj, 513, 733 

%
%

\bibitem[Kundu(2008)]{2008AJ....136.1013K} Kundu, A.\ 2008, \aj, 136, 1013 


%
%
%
\bibitem[Larsen et al.(2001)]{2001AJ....121.2974L} Larsen, S.~S., Brodie, 
J.~P., Huchra, J.~P., Forbes, D.~A., \& Grillmair, C.~J.\ 2001, \aj, 121, 
2974


\bibitem[Lee et al.(2008)]{2008ApJ...682..135L} Lee, M.~G., Park, H.~S., 
Kim, E., Hwang, H.~S., Kim, S.~C., \& Geisler, D.\ 2008, \apj, 682, 135 

%
%
%
\bibitem[McLachlan \& Basford(1988)]{1988MB} McLachlan, G. J., \&
  Basford, K.\ E.\ 1988, Mixture Models: Inference and Application to
  Clustering (New York: M.\ Dekker)

\bibitem[McLaughlin(1999)]{1999AJ....117.2398M} McLaughlin, D.~E.\ 1999,
\aj, 117, 2398

%
\bibitem[Mei et al.(2005a)]{2005ApJS..156..113M} Mei, S., et al.\ 2005a, 
\apjs, 156, 113

\bibitem[Mei et al.(2005b)]{2005ApJ...625..121M} Mei, S., et al.\ 2005, 
\apj, 625, 121 

%
\bibitem[Mieske et al.(2006)]{2006ApJ...653..193M} Mieske, S., et al.\ 
2006, \apj, 653, 193 

\bibitem[Mieske et 
al.(2008)]{2008A&A...487..921M} Mieske, S., et al.\ 2008, \aap, 487, 921 

%
\bibitem[Miller \& Lotz(2007)]{2007ApJ...670.1074M} Miller, B.~W., \& Lotz, J.~M.\ 2007, \apj, 670, 1074 

%
%
%
%
%
\bibitem[Peng et al.(2006a)]{2006ApJ...639...95P} Peng, E.~W., et al.\ 2006a, 
\apj, 639, 95 

\bibitem[Peng et al.(2006b)]{2006ApJ...639..838P} Peng, E.~W., et al.\ 2006b, 
\apj, 639, 838 

\bibitem[Peng et al.(2008)]{2008ApJ...681...197P} Peng, E.~W., et al.\ 2008, 
\apj, 681, 197

\bibitem[Piotto et al.(2007)]{2007ApJ...661L..53P} Piotto, G., et al.\ 
2007, \apjl, 661, L53

\bibitem[Puzia et al.(1999)]{1999AJ....118.2734P} Puzia, T.~H., 
Kissler-Patig, M., Brodie, J.~P., \& Huchra, J.~P.\ 1999, \aj, 118, 2734 

%
%
%
%
%
%
%
%
%
%
\bibitem[Schlegel et al.(1998)]{1998ApJ...500..525S} Schlegel, D.~J., 
Finkbeiner, D.~P., \& Davis, M.\ 1998, \apj, 500, 525

%
%
%
\bibitem[Sirianni et al.(2005)]{2005PASP..117.1049S} Sirianni, M.,
  Jee, M.J., Benítez, N., Blakeslee, J.P., Martel, A.R., Meurer, G.,
  Clampin, M., De Marchi, G., Ford, H.C., Gilliland, R., Hartig, G.F.,
  Illingworth, G.D., Mack, J., \& McCann, W.J. 2005, \pasp, 117, 1049

%
%
%

\bibitem[Spitler et al.(2006)]{2006AJ....132.1593S} Spitler, L.~R., Larsen, 
S.~S., Strader, J., Brodie, J.~P., Forbes, D.~A., 
\& Beasley, M.~A.\ 2006, \aj, 132, 1593 

%
%
\bibitem[Strader et al.(2006)]{2006AJ....132.2333S} Strader, J., Brodie, 
J.~P., Spitler, L., \& Beasley, M.~A.\ 2006, \aj, 132, 2333 (S06)

\bibitem[Strader 
\& Smith(2008)]{2008AJ....136.1828S} Strader, J., \& Smith, G.~H.\ 2008, \aj, 136, 1828 

\bibitem[Tamura et al.(2006)]{2006MNRAS.373..601T} Tamura, N., Sharples, 
R.~M., Arimoto, N., Onodera, M., Ohta, K., 
\& Yamada, Y.\ 2006, \mnras, 373, 601 

%

\bibitem[Tonry et al.(1997)]{1997ApJ...475..399T} Tonry, J.~L., Blakeslee, 
J.~P., Ajhar, E.~A., \& Dressler, A.\ 1997, \apj, 475, 399 

\bibitem[Tonry et al.(2001)]{2001ApJ...546..681T} Tonry, J.~L., Dressler, 
A., Blakeslee, J.~P., Ajhar, E.~A., Fletcher, A.~B., Luppino, G.~A., 
Metzger, M.~R., \& Moore, C.~B.\ 2001, \apj, 546, 681\

%
%
%
%
%
%

\bibitem[Waters et al.(2006)]{2006ApJ...650..885W} Waters, C.~Z., Zepf, 
S.~E., Lauer, T.~R., Baltz, E.~A., \& Silk, J.\ 2006, \apj, 650, 885 


\bibitem[Waters et al.(2009)]{2009ApJ...693..463W} Waters, C.~Z., Zepf, 
S.~E., Lauer, T.~R., \& Baltz, E.~A.\ 2009, \apj, 693, 463 

\bibitem[Wehner et al.(2008)]{2008ApJ...681.1233W} Wehner, E.~M.~H., 
Harris, W.~E., Whitmore, B.~C., Rothberg, B., 
\& Woodley, K.~A.\ 2008, \apj, 681, 1233 

%
%
%
%
%
%
\bibitem[Yoon et al.(2006)]{2006Sci...311.1129Y} Yoon, S.-J., Yi, S.~K., \& 
  Lee, Y.-W.\ 2006, Science, 311, 1129 
%

\end{thebibliography}
\end{document}